# On the lack of women researchers in the Middle East & North Africa

Jamal El-Ouahi[1,2*], Vincent Larivière[3]

[1] *j.el.ouahi@cwts.leidenuniv.nl (Corresponding Author)* - https://orcid.org/0000-0002-3458-7503

Centre for Science and Technology Studies (CWTS), Leiden University, Leiden, Netherlands

[2] Clarivate Analytics, Dubai Internet City, Dubai, United Arab Emirates

[3] *vincent.lariviere@umontreal.ca* - https://orcid.org/0000-0002-2733-0689

École de bibliothéconomie et des sciences de l'information, Université de Montréal, Canada

**Abstract**

Recent gender policies in the Middle East and North Africa (MENA) region have improved legal equality for women with noticeable effects in some countries. The implications of these policies on science, however, is not well-understood. This study applies a bibliometric lens to describe the landscape of gender disparities in scientific research in MENA. Specifically, we examine 1.7 million papers indexed in the Web of Science published by 1.1 million authors from MENA between 2008 and 2020. We used bibliometric indicators to analyse potential disparities between men and women in the share of authors, research productivity, and seniority in authorship. The results show that gender parity is far from being achieved in MENA. Overall, men authors obtain higher representation, research productivity, and seniority. But some countries standout: Tunisia, Lebanon, Turkey, Algeria and Egypt have higher shares or women researchers compared to the rest of MENA countries. The UAE, Qatar, and Jordan have shown progress in terms of women participation in science, but Saudi Arabia lags behind. We find that women are more likely to stop publishing than men and that men publish on average between 11% and 51% more than women, with this gap increasing over time. Finally, men, on average, achieved senior positions in authorship faster than women. Our longitudinal study contributes to a better understanding of gender disparities in science in MENA which is catching up in terms of policy engagement and women representation. However, the results suggest that the effects of the policy changes have yet to materialize into distinct improvement in women's participation and performance in science.

## 1. Introduction

Gender gaps still exist in many areas of society despite the progress made in the past decades. In the Global Gender Gap Report (2021), the World Economic Forum concluded that closing the gender gap will take 135 years from the previous estimated 99 years (2019) due to the impact of the COVID-19 pandemic on several fronts. Gender equality, the fifth Sustainable Development Goal (SDG), is not just a basic human right, but also a prerequisite for a society that is peaceful, wealthy, and sustainable (United Nations, 2015). All United Nations Member States approved the 2030 SDGs in 2015 as a global call to action to enable the well-being of the planet and the people (United Nations, 2015). There has been a rising global interest in directing and analysing state efforts to implement these goals through various approaches (Allen et al., 2018) since the SDGs framework started to be implemented in early 2016.

Scientific research is no stranger to gender gaps. The SDGs clearly identify expanding higher education as a specific goal: "by 2030 ensure equal access for all women and men to affordable and quality technical, vocational and tertiary education, including university" (UNESCO, 2015). Many nation states' higher education institutions expressly target SDGs, which may be regarded as part of the UN-led transformation process. Research institutions are increasingly using SDGs as a framework to influence their missions and translate aspirational goals into quantifiable and sustained benefits. In many countries, educational systems make gender related data publicly available.

However, women's involvement in research from the beginning of their careers has been studied rather thoroughly. The gender indicator can be considered as an indicator for social good: rather than assessing the research "performance" of research institutions, it compares them in terms of a social justice component (Sugimoto & Larivière, May 2019). Many scholars have studied gender gaps in science and sought to explain why women are underrepresented in science (Ceci & Williams, 2007; Williams, 2018). Barriers to education as well as family roles traditionally led by women have negative effects on the balanced development of science systems. These studies look at psychological, sociological, and educational aspects that may have a role in gender disparities.

However, because these studies are primarily concerned with the design of national science systems, they do not provide a global picture of women participation in science. To generate a more holistic picture of women engagement in science across geographies, bibliometrics methodologies have been developed. Such bibliometric methods represent some evidence-based approaches to support the implementation of gender equality and related policies. Halevi (2019) offers an overview of the bibliometric literature on gender differences in science. Research has shown that differences between men and women in science are visible in many different ways, for instance in terms of participation, productivity, collaboration, authorship, research funding, and citations (Fox et al., 2017; Holman et al., 2018; Larivière et al., 2013; Ley & Hamilton, 2008; Lincoln et al., 2012; Shen, 2013). These are critical issues related to gender disparity and bias which must be examined. Women underrepresentation in science in specific regions or countries has been documented for Italy (Abramo, D'Angelo, et al., 2009), France (de Cheveigne, 2009), Québec (Lariviere et al., 2011), and Russia (Paul-Hus et al., 2015). It has also been shown some of the MENA countries perform the worst, globally (Larivière et al., 2013).

Gender differences in first and last authorship have been presented in large-scale cross-sectional studies by Holman et al. (2018) and West et al. (2013). However, in a recent report, UNESCO (2021) provides data that counteracts against stereotypes by revealing that several Arab States have the highest representation of women among engineering graduates, particularly in North Africa with Algeria at 48.5%, Tunisia at 44.2%, and Morocco at 42.2%. Only a few Latin American countries attain comparable numbers, including Peru (47.5%), Uruguay (45.9%), and Cuba (41.7%). This report also denounces the low share of women engineering graduates in the world, especially in several OECD countries such as France (26.1%) the United States (20.4%) and Canada (19.7%). These observations of disparity raise some questions about local science systems. In this study, we aim at developing a more in-depth understanding of the representation of women in science in the wider Middle East and North Africa (MENA) region. Bibliometric analyses at the regional and country level can bring a nuanced portrait of gender differences in science. Such analyses represent logical steps to develop new policies or support existing ones; they are also helpful in assessing the success of policies oriented to improve gender equality in science.

Most of early bibliometric studies provided cross-sectional analyses of gender differences at a specific point in time (Halevi, 2019). Recent research has started to provide longitudinal studies in which the careers of scientists are analysed over time. For instance, Huang et al. (2020) found that while the participation of women in science increased in the past 60 years, gender differences in productivity and impact also increased. Huang et al. (2020) also concluded that career length and drop out rates explain largely the gender differences in productivity and impact. More recently, Boekhout et al. (2021) offered an analysis of the relationship between gender and several factors such as the length of researchers' publication career, the publication productivity as well as the seniority of a researcher. Boekhout et al. (2021) concluded that the difference in the number of men and women starting a career as publishing researchers was the major explanation of gender imbalances among authors of scientific papers.

Our work distinguishes itself from other longitudinal bibliometric studies by focusing on the MENA region in order to better understand the place of women in the scientific research system in this specific region of the world in recent years (2008-2020). Recent gender policies have enhanced legal equality for women in MENA, with noticeable effects in several nations. Through a bibliometric lens, we investigate the status of gender disparities in scientific research in MENA and changes across three cohorts of scholars. Many Middle Eastern and North African economies have made considerable investments in science and technology capacity to promote research and innovation (Schmoch et al., 2016; Shin et al., 2012; Siddiqi et al., 2016), and to move towards knowledge-based economies (OECD, 1996). Several studies analysed the recent growth of scientific production in some of these countries (Cavacini, 2016; Gul et al., 2015; Hassan Al Marzouqi et al., 2019; Sarwar & Hassan, 2015). The following questions are addressed in this research:

1. What are the shares of women authors in the MENA region, by country and field?
2. What is the productivity by gender of authors in the MENA region, by country and field?
3. What is the relationship between gender and author order in the MENA region, by country and field?
4. How has the participation and performance of women in science changed in these countries over time?

The paper is organized as follows. First, we provide a description of the political contest of gender disparities in the region and the engagement of MENA governments with gender policies. Then, we describe the Web of Science data considered in our study and provide details on the disambiguation algorithm and gender inference approach used. Next, we profile countries from MENA based on their share of women authors as well as their productivity across research field. We also analyse the share of women scholars as first and last authors. Finally, we examine the shares of authors per country and per field, the gender differences in productivity by country as well as the gender differences in the first and last authorship. The insights provided by this study are expected to inform science policy makers in the MENA region about gender gaps, thus providing more nuanced interpretations to policies regarding the gender of the scientific workforce in the MENA countries.

## 2. Background

One common argument to explain the differences between countries is that gender gaps narrow through several mechanisms as countries develop. The MENA region is composed of the following countries as defined by the World Bank (October 2019): Algeria, Bahrain, Djibouti, Egypt, Iran, Iraq, Jordan, Kuwait, Lebanon, Libya, Morocco, Oman, Palestine, Qatar, Saudi Arabia, Syria, Tunisia, the United Arab Emirates (UAE) and Yemen. In this study, we also included Pakistan, Afghanistan and Turkey as commonly included in the MENA region (MENAP and MENAT). In line with other studies, we approach the MENA region as a heterogenous group of countries (Hutchings et al., 2010). For each MENA country, Table 1 lists the income group and the 2020 GDP per capita in USD extracted from the World Bank databank[1].

**Table 1. World Bank indicators - Income group and GDP per capita (2020)**

| Country | Population (million) | Income group | GDP per capita in USD (2020) |
|---|---|---|---|
| Qatar | 2.9 | High | 50,124 |
| UAE | 9.9 | High | 36,285 |
| Kuwait | 4.3 | High | 24,812 |
| Bahrain | 1.7 | High | 20,410 |
| Saudi Arabia | 34.8 | High | 20,110 |
| Oman | 5.1 | High | 14,485 |
| Tunisia | 11.8 | Lower middle | 8,536 |
| Lebanon | 6.8 | Upper middle | 4,650 |
| Jordan | 10.2 | Upper middle | 4,283 |
| Iraq | 40.2 | Upper middle | 4,146 |
| Libya | 6.9 | Upper middle | 3,699 |
| Egypt | 102.3 | Lower middle | 3,569 |
| Turkey | 84.3 | Upper middle | 3,522 |
| Djibouti | 1.0 | Lower middle | 3,425 |
| Algeria | 43.9 | Lower middle | 3,307 |
| Palestine | 4.8 | Lower middle | 3,240 |
| Morocco | 36.9 | Lower middle | 3,059 |
| Iran | 84.0 | Lower middle | 2,422 |
| Syria | 17.5 | Low | 1,334 |
| Pakistan | 220.9 | Lower middle | 1,189 |
| Yemen | 29.8 | Low | 758 |
| Afghanistan | 39.0 | Lower middle | 517 |

---

[1] https://databank.worldbank.org/source/world-development-indicators

The Gulf Cooperation Council (GCC) countries are all part of the High-Income group with a substantially higher GDP per capita than other MENA nations, whereas most North African and Middle Eastern countries are lower income countries. Jayachandran (2015) argues that although the development process can account for most of the GDP/gender disparity relationship across the whole labour force, specific society factors also play a role.

In her study, Jayachandran (2015) explores several root causes of gender inequality in poor countries: economic underdevelopment, cultural factors and persistence of gender norms when economic conditions evolve. But Narasimhan (2021) found that, as per capita wealth rises, the proportion of women in science rises at first, then declines. This is in stark contrast to the well-established U-shaped curve for women's involvement in the labour market as a whole, implying that there are variables in science culture that result in opposite tendencies to those seen in the broader population. As a result, women make up a significantly larger proportion of the scientific workforce in many developing nations than in economically developed ones (Narasimhan, 2021). Narasimhan (2021) also found different patterns in terms of women scholars' retention in developing and developed countries.

The overall low women labour participation in MENA is also often attributed to oil (World Bank, 2012). The economic structure, social norms, and institutional characteristics of oil-rich economies have been claimed to hinder women from working in the formal sector (Moghadam, 2005). Ross (2008) argued this is primarily due to the impact of oil on these countries' economies, as well as a lack of employment incentives for women due to significant subsidies and family benefits which encourage women to stay at home. More recently, Kucuk (2013) provided an empirical study data that questioned the notion that religion and oil are to blame for gender inequality in MENA, Arab, and Muslim majority nations. Kucuk (2013) concluded neither of these reasons explains the inequalities completely.

Another argument is the turmoil experienced by certain MENA nations during the so-called 'Arab Spring,' with social unrest, civil conflict, or armed insurrection. There has also been some debate about the effects of the so-called 'Arab Spring' on scientific capacity in Arab countries (Ibrahim, 2018; Turki et al., 2019): many socio-political factors come into play but the armed rebellion or civil war in Syria, Libya and Yemen have certainly impacted the national science systems of these countries.

In recent years, many other MENA economies have made significant social progress in terms of human development (World Bank, 2012). While women have lower educational levels than men in the overall population, MENA countries perform relatively well in international comparisons when it comes to the younger generation (OECD, 2014). In many MENA nations, gender disparities in education have been virtually erased, and women currently predominate in tertiary education. However, this situation varies greatly across the region. Scores are still low in low-income countries (OECD, 2014). And the overall gender equality we observe in tertiary education does not necessarily reflect in academic research in most countries.

Several studies have looked at other reasons. Some analysts believe the mismatch is also due to supply-side factors, since young women have increasingly entered the market at a period when employment prospects for both men and women have remained stagnant (Assaad et al., 2020). According to the World Bank, low rates of women's engagement in the public sphere are due to conservative gender norms, legal and institutional impediments, and incentives and opportunities produced by local economic systems in MENA (World Bank, 2013).

With the Women, Business and the Law (WBL) project[2], the World Bank measures legal differences in men's and women's access to economic opportunities in 190 countries. Thirty-five parts of the law are evaluated using eight indicators, each with four or five binary questions. Parenthood, Workplace, Pay, Marriage, Mobility, and Entrepreneurship are all indicators that represent different stages of a woman's career. Calculating the unweighted average of the questions within an indicator and scaling the result to 100 yields indicator-level scores. The average of each indicator is then used to produce overall scores, with 100 as the highest possible score. These indicators provide objective and quantitative benchmarks for worldwide gender equality advancement. This data is also useful for research and policy discussions on enhancing women's economic opportunities.

Figure 1 presents the WBL index performance in 2008 and 2021 by region: High income OECD (HIOECD), Europe-Central Asia (EUCA), Latin America and Caribbean (LATAM), East Asia & Pacific (APAC), Sub Saharan Africa (SSA), South Asia (SA), and MENA. The WBL index increased for all regions. Although the MENA region is catching up with other emerging regions, MENA still lags behind all other regions of the world, with women in the region having about half the rights of men across all the career stages evaluated. This suggests there is still room to improve the legal framework in MENA.

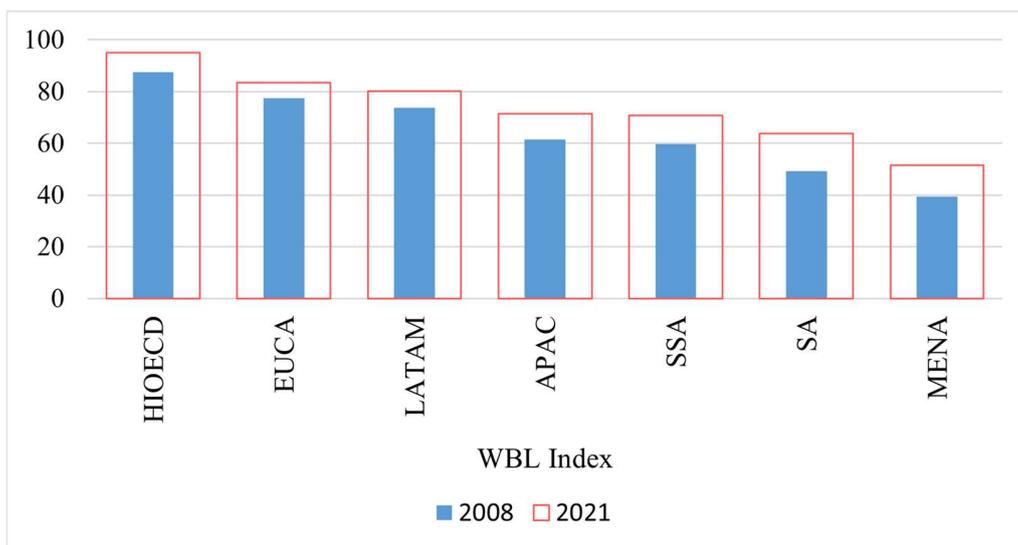

**Figure 1. WBL index performance by region (2008 and 2021)**

---

[2] https://wbl.worldbank.org

As mentioned earlier, although the MENA region is lagging compared to other regions of the world, its WBL index increased between 2008 and 2021. MENA nations have recently adopted new gender-transformative policies that address the root causes of gender inequity: previous policies have sought to fit women into inequitable systems, but new ones aim at fixing the system to provide employment for women and reduce gender disparities. Figure 2 shows the WBL index performance by MENA nation in 2008 and 2021. As per the World Bank, the WBL index increased for most countries in MENA, with Turkey, the UAE, Saudi Arabia, Morocco and Tunisia leading the region. Starting from a relatively low score, UAE, Jordan and Saudi Arabia show a large increase of their WBL index between 2008 and 2021.

In recent years, the UAE government has made gender equality and women's economic empowerment a primary policy focus. The country implemented historic reforms led by the UAE's Gender Balance Council to boost women's economic empowerment, such as prohibiting gender-based discrimination in employment and removing job restrictions imposed on women in several sectors. Despite the COVID-19 pandemic situation, a landmark reform package was passed. For the first time in the MENA region, this reform included fully paid parental leave to both women and men employees in the private sector. The Labour Law was also amended to require equal compensation for equal work across different industries and sectors.

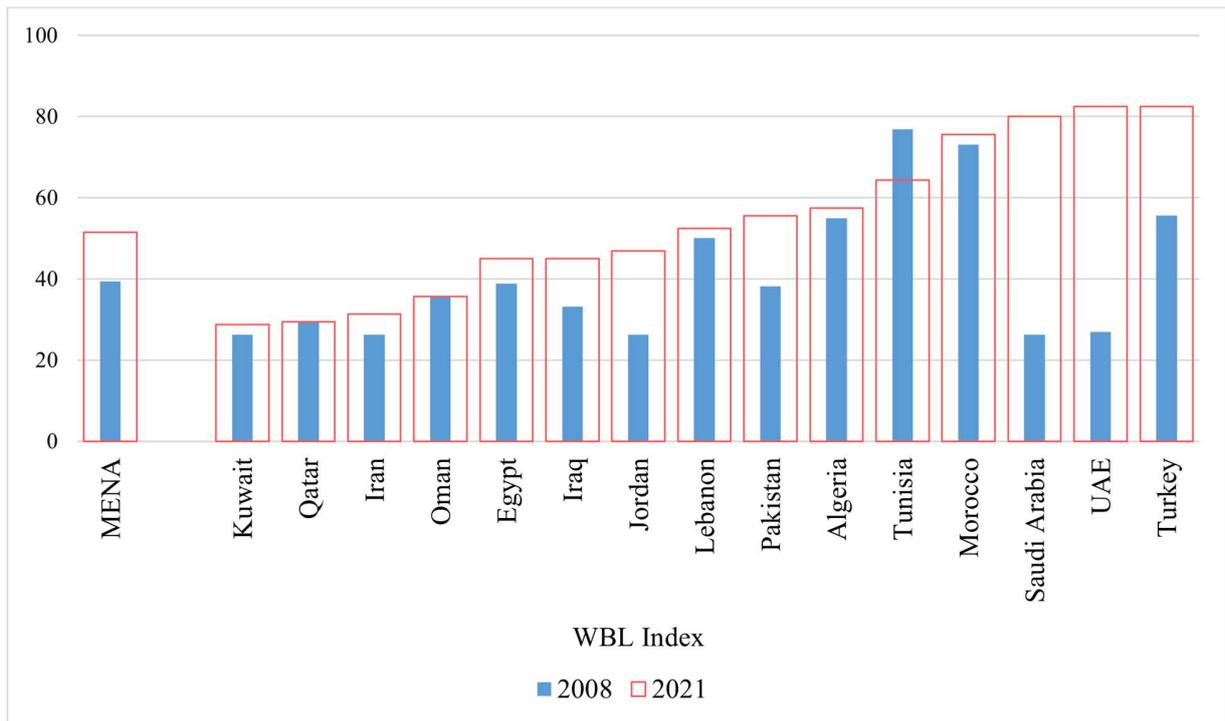

**Figure 2. WBL index Performance by MENA country (2008 and 2021)**

The World Bank's Women, Business and the Law group reported these improvements in the UAE's overall WBL score, with an increase from 30 out of 100 points in 2019 to 82.5 out of 100 points in 2021.

The United Arab Emirates' reforms were preceded by Saudi Arabia's ground-breaking changes. Saudi Arabia passed landmark legislation in July 2019 that changed women's status in society and gave them more freedoms. The reforms in Saudi Arabia include a variety of measures aimed at easing constraints on women's right to free movement both within and outside the country. A set of laws also increased women's protections at work and in their access to financial services. Employers are no longer allowed to discriminate based on gender in hiring and advertising jobs, and financial institutions are no longer allowed to discriminate based on gender when providing financial services.

The reforms in the UAE and Saudi Arabia have improved legal equality for women in recent years. Such reforms will most likely continue to boost the involvement of women in the respective economies. Changes led locally by Saudi Arabia and the UAE leaders are some of the keys to achieving positive outcomes and increasing women's economic involvement. This is critical not only for motivating changes, but also for sharing reform experiences, success factors, and lessons learned.

The ripple effect of the UAE and Saudi Arabia changes is even more apparent in other nations. For example, there are certain movements toward equality, human rights, and justice in Jordan, as well as duties to enact equality legislation. Jordan's government has undertaken reforms in the areas of flexible work arrangements, employer-provided childcare, and the lifting of limitations on women working in some sectors and at certain hours since 2017. Despite considerable initiatives and regulatory and legislative reforms to address gender equality as serious actions by the government and policymakers, women are still under-represented in the workplace in Jordan in contrast to men (Koburtay et al., 2020).

Recently, the Human Rights Watch (2021) reported that discriminatory restrictions affect women's independence to marry, study, work and travel. The Qatar Government (2021) responded that this report was inaccurate and was not aligned with Qatar's constitution, laws, policies. In its response, Qatar's government listed a few examples such as its investment to ensure all women have access to education and opportunities across all sectors, especially in STEM. It also mentioned the progress made with the high percentage of women enrolled in university programs. Finally, continuity in enforcing, introducing, and expanding policies that provide freedom to women to make their own decisions was highlighted.

When it comes to reforming laws for greater women's empowerment in the MENA region, countries stand at different levels of engagement with gender equality policies. Recent sets of reforms in some countries will boost the prioritization of legislation to reduce gender inequalities in the region. While these reforms are anticipated to have a long-term influence on women's participation in MENA, enforcement mechanisms will also be required to help relevant government institutions accelerate their implementation.

## 3. Data and methods

### 3.1. *Data*

Data for this study is extracted from four Web of Science citation indices (the Science Citation Index Expanded, the Social Sciences Citation Index, the Arts & Humanities Citation Index, and the Conference Proceedings Citation Index). Linking between authors and their affiliations in Web of Science began in 2008. Included in this study were all items published between 2008 and 2020 with at least one institutional address located in the MENA region.

To identify a set of publications with a given author during the study period, we used the disambiguation algorithm proposed by Caron and van Eck (2014) which produces the highest quality results as demonstrated by Tekles and Bornmann (2019). Caron and van Eck (2014) reported an average precision of ~0.97 and a recall of ~0.91 on their test data set. The dataset under study was comprised of 1,656,510 publications contributed by 1,124,256 disambiguated authors. Since authors with one or two publications are more likely to be artefacts of the disambiguation algorithm, we considered only authors with at least three publications. On the whole dataset, 340,608 disambiguated authors satisfy this criterion and 783,648 were excluded. In our paper, we use the different words 'researcher', 'scholar' and 'author' to refer to disambiguated authors.

### 3.2. *Gender inference*

To infer a gender for authors, we used the same algorithm as in the 2021 edition of the Leiden Ranking based on first name and country of affiliation. The algorithm is built on the following approach. First, we assign countries to each author, based on their affiliation on research papers. If the country of the author in his or her first publication is the same as the country the author is most often associated with in his or her set of papers, we then consider this country as the author's country of origin. Then, we used three tools to infer a gender: Gender API (https://gender-api.com), Gender Guesser (https://pypi.org/project/gender-guesser), and Genderize.io (https://genderize.io). It has been shown Gender API performs better as evaluated in a previous study (Santamaría & Mihaljević, 2018). The first name of the author combined with a country of origin were provided as inputs to these tools. Overall, men accounted for 49% of the researchers, and women for 29% of the researchers. The remaining 22% could not be identified either way and were excluded from our study. We used a dataset created by Larivière et al. (2013) to evaluate the accuracy of our gender inference. We discuss the results of this evaluation in Appendix A. For authors inferred to be men, the approach of Larivière et al. (2013) is slightly more accurate than ours. However, for authors inferred to be women, their approach has a lower accuracy than ours.

### 3.3. *Field assignment*

We also assigned scientific disciplines to authors. In the Web of Science, each scientific journal belongs to one or multiple subject categories. We used the OECD's Fields of Science and Technology (FOS) classification. The FOS classification has two hierarchically levels: 6 major codes and 42 minor codes. We used only the six major OECD fields: Agricultural Sciences (AGR),

Engineering & Technology (ENG), Humanities (HUM), Medical & Health Sciences (MED), Natural Sciences (NAT) and Social Sciences (SOC). The Frascati Manual 2002 recommends that the major fields of science and technology should be adopted as the functional fields of a science classification system. It also recommends that this classification should be used for the R&D expenditure of the governments, higher education and Private/Non-Private sectors. We mapped the publications' WOS classifications to the FOS classification based on the correspondence provided by Clarivate Analytics (2012). Some authors were linked to multiple disciplines since a source may belong to more than one discipline.

### 3.4. Cohort assignment

For each researcher, we considered the year of first publication as the start their research career. We defined three fixed populations of authors from MENA: *2008-cohort, 2012-cohort* and *2016-cohort,* for authors who wrote their first paper in 2008, 2012 and 2016 (respectively) and published at least three papers in their career-to-date. All publications of these authors are considered, including the publications without a MENA affiliation for the authors who moved outside MENA. Table 2 lists the number of women and men authors for each cohort.

**Table 2. Number and share of women and men authors for each cohort**

| *Gender* | *2008-cohort* | *2012-cohort* | *2016-cohort* |
|---|---|---|---|
| Women | 5,178 (32%) | 6,968 (36%) | 7,777 (39%) |
| Men | 10,854 (68%) | 12,472 (64%) | 12,114 (61%) |

### 3.5. Authorship productivity

We define the productivity as the average number of papers per author per year. We calculated their productivity during the study period. Here we used a full counting approach. All co-authors of a publication are considered to have contributed to the publication.

### 3.6. Authorship position

To address seniority and leadership in scientific output, we focused specifically on the first and last author of a publication. In most fields, when a researcher is the first or last author on a publication, it is more likely that the author had a central role in the research project in terms of execution, guidance, management or funding (González-Alcaide et al., 2017; Henriksen, 2019). The first author of a publication often represents the most important contributor (Larivière et al., 2016), that is the one who conducted the majority of the authorship tasks. This is typically a more junior scholar (Larivière et al., 2016) and is increasingly associated with more than one individual (Lapidow & Scudder, 2019). Last author, particularly in the natural and biomedical sciences, is largely associated with the senior author who provided resources and mentorship in the project (Larivière et al., 2016). Not all fields adhere to this structure of authorship: e.g., Mathematics, Economics, and High Energy Physics still largely employ alphabetical ordering (Waltman, 2012). However, less than 4% of articles globally intentionally use alphabetical listing (Waltman, 2012).

Therefore, we used the first and last authorship as a proxy for leadership or seniority in research. On authorship, we used the assumption that the first-listed author has the greatest share of responsibility for the publication. It is worth reminding there is no universally accepted solution for allocating author credit for research publications in bibliometric analysis. Gender differences in first and last authorship have been presented in large-scale cross-sectional studies by Holman et al. (2018) and West et al. (2013).

The approach to calculate the probability of being first of last author of a publication for men and women in a specific year of their publishing career is explained as follows: Let us consider the authors of X-cohort (researchers who started their publishing career in year X). We first identify all the authorships in year Y of the men and women authors who started their career in X. For explanation purposes, such a set of authorships is represented in Table 3 with 5 publications and 5 authors each. Women authors are marked in orange and men authors are marked in blue. Only the authors who started their career in X are marked as bold/underlined and are considered in our calculations, (W1; W3; W4; W5) and (M2; M3; M7; M8; M9).

Table 3. Fictional set of publications contributed by the X-cohort in year Y

| Paper | Author 1 | Author 2 | Author 3 | Author 4 | Author 5 |
|---|---|---|---|---|---|
| P1 | **W1** | M1 | **M2** | W2 | **M3** |
| P2 | **M2** | M4 | **W3** | **M3** | M1 |
| P3 | M5 | **W4** | M6 | M1 | **M7** |
| P4 | **W5** | **M8** | W6 | **M2** | **W1** |
| P5 | M1 | **W4** | M5 | **M9** | **M2** |

The probability of being first/last author for a specific gender is calculated by dividing the number of first/last authorships of that specific gender by the number of authorships of the same gender from the same cohort. In the case of the publications shown in Table 2, the probabilities are listed in Table 4 along with their calculations:

Table 4. Probabilities of being First or Last author for the X-cohort based on the fictional set of authorships in year Y.

| Probability | First Author | Last Author |
|---|---|---|
| **Women** | 0.33 (2/6) | 0.16 (1/6) |
| **Men** | 0.11 (1/9) | 0.33 (3/9) |

This approach answers the following question: "For all women (men) researchers that started in year X, when someone published a paper in Y, what was the probability they were the first (last) author of the paper?"

*3.7. Limitations*

The bibliometric data and methods used have well-documented limitations. The first limitation is related to the author disambiguation algorithm, which may split a single author into several authors when they have unconventional publication behaviour. Therefore, we may undercount the productivity of individuals who exhibit behaviours that lead to splitting: e.g., changing fields, institutions, or names. Gender disambiguation also introduces errors (incorrect gender assignment and restriction to binary classification (Kaitlin et al., 2019)) and data loss (e.g., 22% of researchers in our dataset who could not be classed). The limitations of algorithmic approaches to gender inference have also been discussed by Mihaljević et al. (2019). Finally, for each author, the year of the first publication is used as a proxy of the year in which the author entered the science system. Considering the time required to publish a paper, such author has most probably entered the research system earlier. Finally, the indicators we use as leadership or seniority have also limitations, as discussed above (Glänzel et al., 2016; Waltman, 2016). Without specific descriptions of the role of each author, such as promoted in the Contributor Roles Taxonomy (CRediT) initiative, it remains difficult to assess authors' contributions (Larivière et al., 2021). Finally, we studied only gender gap differences on few aspects of scientific research based on Web of Science Core Collection data. Since national languages are often generally used in fields where we have national applications, data from other citation indices with more regional content, such as the Arabic Citation Index, might also be used in future work to provide more comprehensive analysis for policy makers. The results should be interpreted in light of these limitations.

## 4. Results

*4.1. Share of women authors by country for each cohort*

First, we analysed the shares of researchers for each country in MENA during the study period. Figure 3 presents the shares of women researchers for each cohort by country. It is worth mentioning that Afghanistan, Bahrain, Djibouti, Libya, Palestine, Syria and Yemen have on average less than 15 researchers across all cohorts. Due to the small presence of active researchers in these countries, these countries were excluded from our dataset to draw reliable statistics for the region.

As a region, MENA has a lower share of women authors compared to the global share of women authors for the three cohorts. However, some interesting cases stand out. Tunisia and Lebanon have the highest average shares of women authors (respectively 58% and 48%) and are the only countries with an average share of women scientists above 40% for the three cohorts. Turkey, Algeria, and Egypt follow with an average share of women scholars that is higher or equal than the MENA average (36%). Iraq, Saudi Arabia, Oman, Pakistan and UAE have the lowest average shares, respectively 14%, 21%, 26%, 27% and 27%.

Overall, shares of women authors have increased on average by 7% when comparing the 2008 and 2016 cohorts. Tunisia, Qatar, Jordan and Iran have shown the highest increases, respectively 19%, 18%, 17%, and 10%. However, for the same comparison, the share of women scholars in Iraq decreased by 1%, and it grew only by 1% in Saudi Arabia, Kuwait and Egypt. Although the increase of the share of women publishing scientists is notable in several MENA countries, these shares remain lower than the shares in the MENA region and the world.

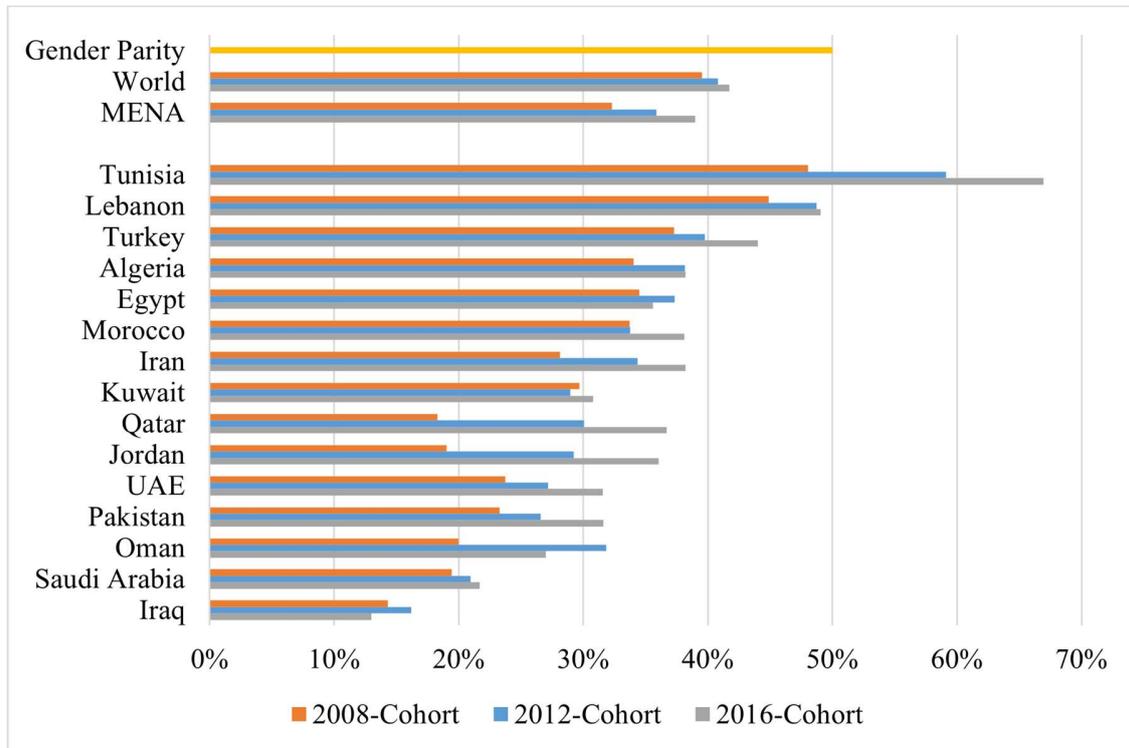

**Figure 3. Share of women scholars per gender and country for each cohort**

## 4.2. *Share of women authors by field*

We analysed the proportions of women authors with their corresponding OECD field for each cohort. Figure 4 presents these shares. We clearly see there are relatively more men authors than women authors across all fields. However, we notice some differences. In the Medical & Health Sciences, the share of women authors is the highest with an average of 42% for the three cohorts, followed by Humanities (40%). Women scientists represent about 36% of the scientists in Agricultural Sciences, Social Sciences, and Natural Sciences. Finally, Engineering & Technology has the lowest average share of women authors with 31% for the three cohorts. In terms of trends in MENA, the shares of women scholars have increased in all the fields, except in the Humanities.

The low proportion of women scientists in MENA could be due to their overall lower representation in the whole region as shown previously. This contrasts with the fact that women exceed men in tertiary education in half of MENA countries, and more women graduate in science, technology, engineering, and mathematics than in many OECD countries (OECD, 2013). Shalaby (2014) refers to this situation as the "Paradox of Women Economic Participation in the Middle East and North Africa" region.

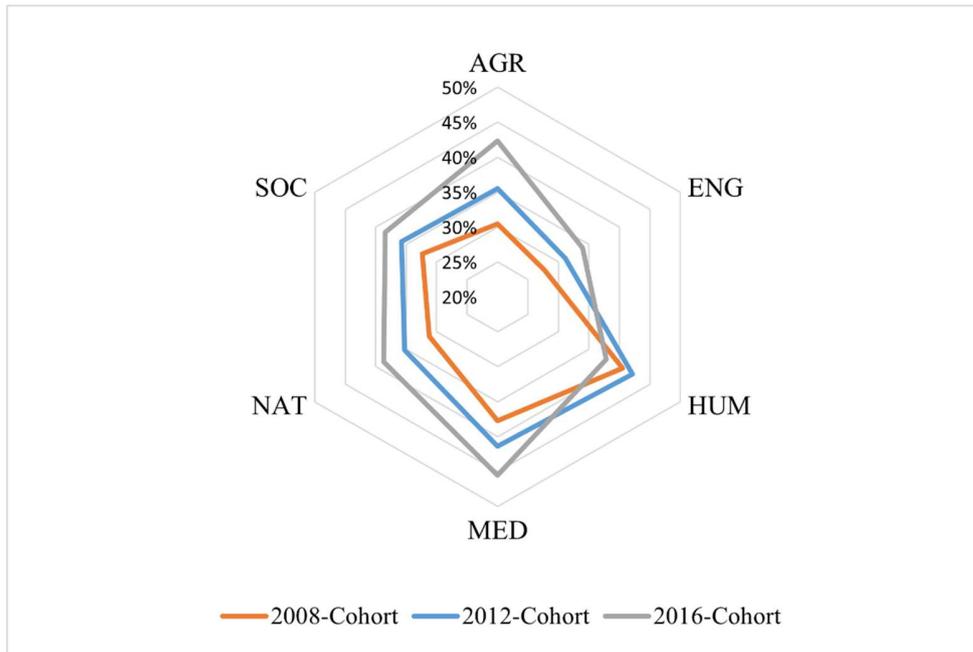

**Figure 4. Share of women scholars for each cohort by OECD field in MENA**

## 4.3. *Gender differences in productivity*

The productivity of men and women authors was also examined for each gender. Figure 5 shows the average number of publications to which each cohort has contributed.

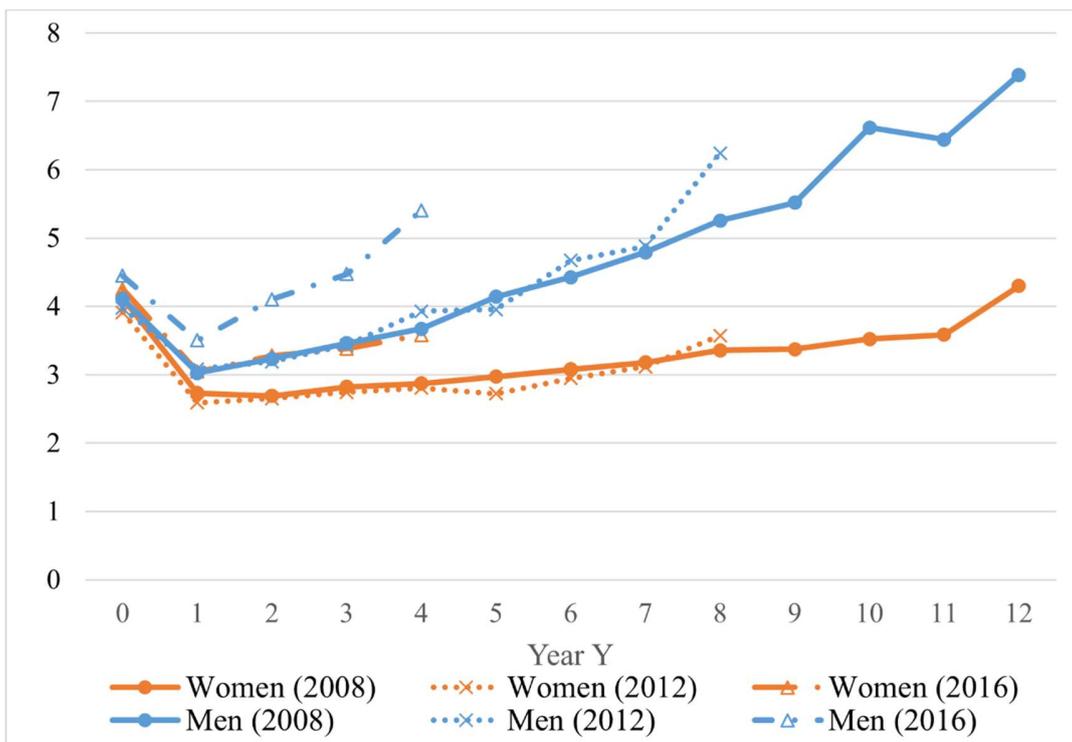

**Figure 5. Average number of publications by cohort and year of publishing career Y**

Figure 5 shows that, overall, men who entered the science system in 2008, 2012 or 2016 had a higher productivity than women who entered the science systems in the same years. There was no substantial difference in terms of productivity between women and men in their first year of their career for each cohort. However, in the following years, the productivity of men was between 11% and 51% higher than the productivity of women with this gap increasing over time. Also, we notice the productivity of the 2016-cohort has increased for both genders when compared with the productivity of the 2008 and 2012 cohorts in the same career years, but this increase was higher for men than women.

The low productivity of women scientists in MENA could be due to their overall lower representation in the whole region as mentioned earlier. Another reason is that women appear to be slightly less likely than men to pursue a career as a publishing researcher, but the difference is small (Boekhout et al., 2021). There are also well-known differences between disciplines in the average number of publications per researcher. In some disciplines, men are overrepresented with a larger average number of papers per researcher than women. In other disciplines, women are overrepresented with a smaller number of publications per researcher. We have broken down our analysis by OECD field and we calculated the average number of publications by women and men authors in year 5 of their career for each cohort by using the full counting method. This breakdown is represented in Figure 6.

The overall gender difference in productivity shown in Figure 5 can also be observed at the field level. Men who entered the research system in 2008, 2012 and 2016 had a higher productivity in year 5 of their career than their women colleagues who entered the science system in the same year across all the 6 OECD fields. However, the relative differences between men and women at the field level are smaller in Social Sciences and Humanities than at the overall level. But we observe a widening gap in Social Sciences for the 2016-Cohort. The overall gender difference in productivity is partly due to a relative overrepresentation of men in certain fields with a larger average number of publications per author. Engineering & Technology, and Natural Sciences have the highest average number of publications per researcher. These fields have also a large overrepresentation of men among researchers entering the science system.

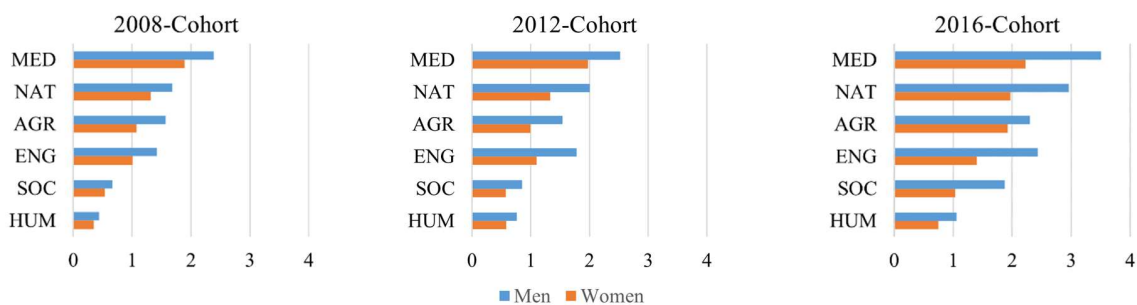

**Figure 6. Average number of publications by OECD field by women and men authors in year 5 of their career for each cohort**

Several studies have documented the productivity differences between men and women researchers (Halevi, 2019). Men researchers produce on average more publications than their women peers. Mishra et al. (2018) also observed these differences in a large-scale study and they found that differences in productivity cannot be explained by differences in career lengths. However, Huang et al. (2020) found that differences in productivity are largely explained by career length and dropout rates, concluding that "men and women publish at a comparable annual rate". They found that the "number of publications per year for women and men authors are largely indistinguishable" and refer to as a 'gender invariant'. Similarly to what Boekhout et al. (2021) found, our results are in contrast with these results. Perhaps, the different findings could be explained by the older time period covered by Huang et al. (2020). Furthermore, our analysis focuses on a specific region and a specific period in the career of the population under study whereas Huang et al. (2020) analysed the average productivity during entire careers.

### 4.4. *Gender differences in lead authorship*

Here, we focus on the three cohorts, i.e. the researchers from MENA who entered the research system in 2008, 2012 and 2016 and produced at least 3 papers during the study period. For these groups of authors, we analysed the time trends in the probabilities of being first or last author of a publication during the study period. We calculated these probabilities for each year as explained earlier in the Data and Methods section. We represent these time trends in Figure 7.

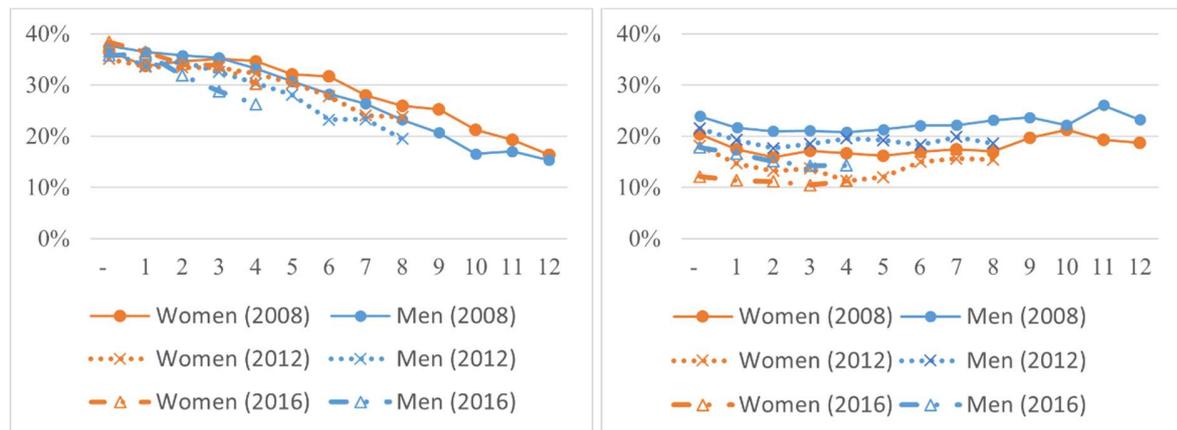

**Figure 7. Probability of being first (left chart) or last (right chart) author of a publication for men and women researchers from MENA who started their publication career in 2008, 2012 and 2016.**

The probability of being the first author of a publication decreased over time for all cohorts. Researchers were more likely to be first authors of a publication in the early years of their careers than in the latter years. We also note some gender differences in terms of authorship position. The probability of being first author is slightly higher for men than for women until the fifth year of their publishing career. As shown in the left chart, men in MENA were more likely to be first author of a publication than their women colleagues in earlier years of their career. However, after the fifth year, women were more likely to be first author. When comparing the three cohorts, we do not notice a shift in terms of probability of being first author.

Furthermore, as shown in the right chart, men were more likely to be last authors than women in early years of their careers as well as in the later years for all cohorts. The probability of being last author was between 1.04 and 1.73 times higher for men than for women. This probability decreased slightly for researchers in the first 3 to 5 years of their career. Then, it increased earlier for men than for women of the same cohorts. As mentioned earlier, we can use the last authorship position as a proxy of seniority. This suggests that, on average, men reached a more senior position faster than women. Boekhout et al. (2021) have shown that the gender differences in the probability of continuing a career as publishing researcher are small and they play no role in explaining gender disparities in the scientific literature. Robinson-Garcia et al. (2020) find that women are less likely to serve in a leading author role early in their career and this has ramifications for attrition.

If we compare the three cohorts, we also observe a change in terms of likelihood of being last author. For example, researchers of the *2008-cohort* were more likely to be last authors than authors of the *2012-cohort* in the first years of their publishing careers. Similarly, researchers of the *2012-cohort* had a higher probability of being last author than researchers of the *2016-cohort* in the early years of their career. Additionally, the *2008-cohort* (men) and the *2012-cohort* (men) have reached the same level of probability 4 years after the start of their publishing career, whereas it took 8 years for the *2012-cohort* (women) to reach the same level of probability of the *2008-cohort* (women). However, we notice the *2016-cohort* (women) reached the same level of probability of the 2012-cohort (women) in 5 years, while we can observe a constant gap for men researchers of the same cohorts.

Different disciplines have different norms in authorship (Biagioli & Galison, 2014). Therefore, we calculated the probability of being last or first author of a publication after five years of activity by discipline for the same groups of researchers who started their publication career in 2008, 2012 or 2016. These probabilities are represented in Figure 8. For all cohorts, in Humanities, the men researchers had a higher probability than their women colleagues to be first author of a publication in the fifth year of their career. This difference in terms of probability of first authorship decreased when comparing the three cohorts. In Agricultural Science and Engineering & Technology, women were more likely to be first author of publications after five years of activity for the three cohorts as well. In Natural and Medical Sciences, there is no substantial difference. Finally, in Social Sciences, we do not notice any gender difference in terms of first authorship for the 2008-cohort. However, the 2012 and 2016 cohorts show a difference in this particular field with men active researchers more likely to be first authors than their women colleagues.

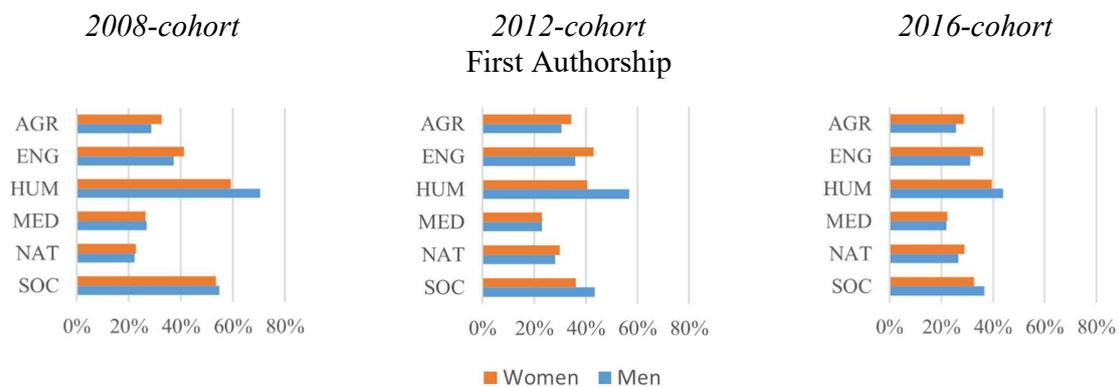

*2008-cohort*　　　*2012-cohort*　　　*2016-cohort*
First Authorship

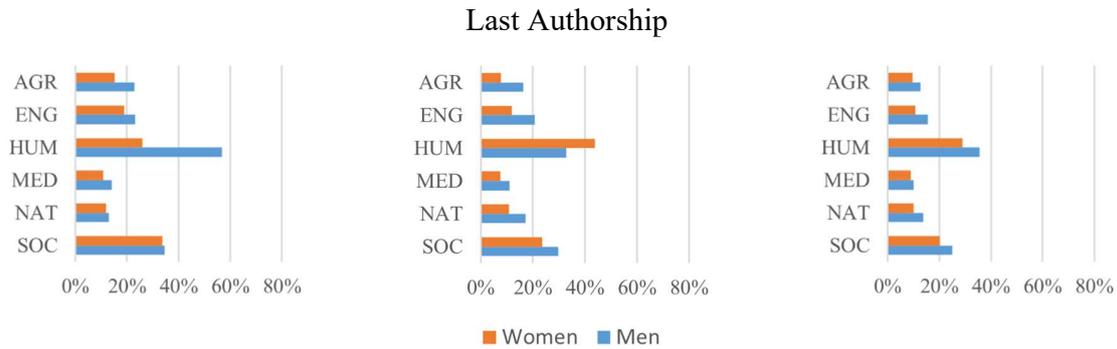

Last Authorship

**Figure 8. Probability of being first author (top) or last (bottom) by OECD field in year 5 of their career for men and women who started their career in 2008, 2012 and 2016.**

We notice that the women researchers of the three cohorts are more likely to be first authors in the fifth year of their career than men researchers in Agricultural Sciences and Engineering & Technology. There is no substantial difference in Medical Sciences and Natural Sciences. However, men researchers of all cohorts have a higher probability of being last authors than their women peers after 5 years of activities across all fields, except in Humanities for the 2012-cohort. These results tend also to confirm the division between *feminized* areas of research (Humanities and Social Sciences) and men dominated fields (Natural Sciences, Engineering and Technology) (Alon & Diprete, 2015; Trusz, 2020).

*4.5.* *Country profiles*

In this section, we analyse the proportion of women scholars by country as well as the gender differences in productivity. We also study the first and last authorship position by gender.

Figure 9 presents the shares of women scientists for each cohort per country and OECD field. This heatmap allows us to clearly see which countries have attained or are close to gender parity in specific fields. Countries are sorted from the lowest to the highest share of women authors at the country level. It is worth reminding that some authors might have been active in several OECD fields based on the multiple categories of the journals where they have published their papers.

|      |     | Iraq | Saudi Arabia | Pakistan | Qatar | Jordan | Oman | UAE | Kuwait | Iran | Morocco | Egypt | Algeria | Turkey | Lebanon | Tunisia |
|------|-----|------|--------------|----------|-------|--------|------|-----|--------|------|---------|-------|---------|--------|---------|---------|
| 2008 | AGR | -    | 12           | 22       | 40    | 18     | 10   | 38  | 33     | 21   | 36      | 38    | 35      | 37     | 29      | 56      |
|      | ENG | 18   | 16           | 19       | 13    | 13     | 16   | 19  | 23     | 23   | 27      | 31    | 29      | 34     | 34      | 46      |
|      | HUM |      | 11           | 14       | -     | 13     | 50   | 20  | 20     | 34   | 20      | 50    | 100     | 50     | 45      | 36      |
|      | MED | 6    | 22           | 27       | 24    | 27     | 29   | 29  | 32     | 35   | 46      | 41    | 46      | 40     | 48      | 58      |
|      | NAT | 19   | 18           | 23       | 16    | 16     | 20   | 23  | 28     | 27   | 29      | 35    | 35      | 36     | 41      | 49      |
|      | SOC |      | 22           | 18       | 11    | 27     | 6    | 22  | 38     | 24   | 25      | 28    | 33      | 42     | 52      | 40      |
| 2012 | AGR | 33   | 22           | 24       | -     | 31     | 44   | 50  | -      | 29   | 36      | 33    | 38      | 45     | 65      | 71      |
|      | ENG | 15   | 18           | 22       | 27    | 23     | 33   | 23  | 23     | 29   | 26      | 32    | 34      | 36     | 36      | 56      |
|      | HUM | 33   | 21           | 13       | 18    | 25     |      | 22  | -      | 35   | 67      | 39    | 43      | 52     | 60      | 100     |
|      | MED | 16   | 28           | 33       | 33    | 35     | 33   | 35  | 33     | 42   | 51      | 44    | 60      | 41     | 58      | 68      |
|      | NAT | 16   | 21           | 25       | 33    | 29     | 38   | 27  | 32     | 34   | 30      | 38    | 38      | 40     | 45      | 58      |
|      | SOC | 40   | 15           | 18       | 39    | 39     | 38   | 30  | 31     | 32   | 32      | 43    | 26      | 44     | 53      | 48      |
| 2016 | AGR | 21   | 13           | 35       | 20    | 31     | 29   | 13  | -      | 37   | 47      | 36    | 51      | 49     | 65      | 80      |
|      | ENG | 12   | 16           | 27       | 29    | 29     | 14   | 31  | 22     | 32   | 37      | 32    | 33      | 38     | 38      | 70      |
|      | HUM | -    | 44           | 35       | 29    | 17     | -    | 23  | 100    | 31   | -       | 21    | 33      | 45     | 67      | 100     |
|      | MED | 15   | 25           | 38       | 43    | 47     | 47   | 39  | 31     | 48   | 45      | 41    | 57      | 49     | 54      | 72      |
|      | NAT | 13   | 21           | 31       | 37    | 31     | 27   | 32  | 29     | 38   | 37      | 36    | 39      | 44     | 51      | 68      |
|      | SOC | 10   | 23           | 22       | 34    | 22     | 12   | 20  | 31     | 35   | 37      | 34    | 41      | 49     | 70      | 57      |

**Figure 9. Percentage of women scholars per country and OECD field (2008-2020)**

Tunisia has overachieved gender parity in terms of number of scholars across all fields. Lebanon, Turkey, Algeria, Bahrain, Egypt, and Morocco follow. Some countries such as Iraq, Saudi Arabia, Pakistan, Qatar have low shares of women scientists in most fields. This heat map allows us to see how the shares of women scholars per country and per field have evolved when comparing the three cohorts. For most countries in the MENA region, we notice an increase of these shares over time across all fields.

*4.5.a. Gender differences in productivity by country*

Figure 8 shows the average number of publications for women and men researchers who started their career respectively in 2008, in 2012 or in 2018 in year 5 of their career. As previously mentioned, we also use a full counting approach here, i.e., each author of a publication is considered to have produced a full publication.

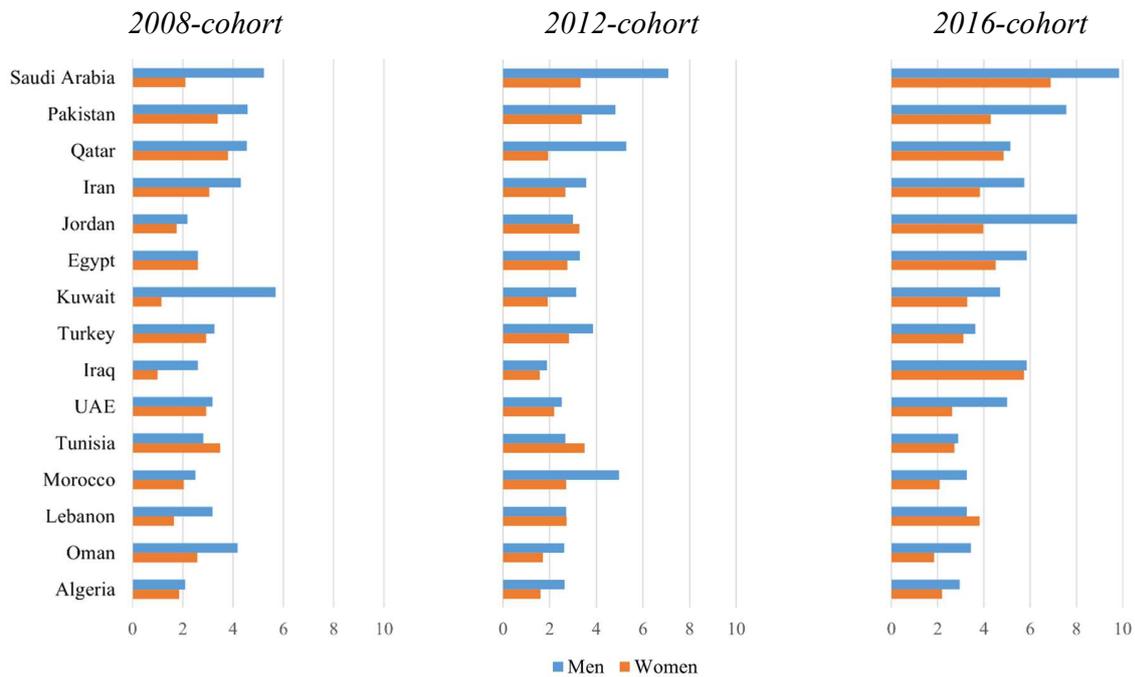

**Figure 10. Average number of publications of researchers by gender in year 5 of their publishing career for each cohort**

This figure shows that, overall, men have a higher publication productivity than women in most MENA countries for all three cohorts. However, few countries standout such as Tunisia (2008 and 2012), Lebanon (2012 and 2016), Jordan (2012), Qatar and Iraq (2016). Based on the statistics of each cohort, we can also analyse how the gender differences in terms of productivity have evolved over time. First, we notice the overall average number of publications in year 5 of the career of the researchers has increased for both genders and for all countries with an average of 3 papers for the 2008-cohort, 3.1 papers for the 2012-cohort and 4.4 papers for the 2016-cohort. Then, we also notice a widening gap for many countries.

As mentioned earlier, there are well documented differences in the average number of a publications of a researcher between disciplines (Halevi, 2019). The gender differences we observe in Figure 10 might be due to an underrepresentation of women in disciplines with a larger average number of publications per researcher as noted in our previous analysis on gender differences in productivity in MENA. These differences can also be partially due to women having more gaps in their publication career than men. For example, women researchers might have to pause their research careers for family reasons such as pregnancy and/or maternity leave.

*4.5.b. First and last authorship by gender and country*

In this section, we analyse the probability of being first or last author of a publication for the three cohorts after 5 years of publishing activity. These probabilities are represented in Figure 11. As shown earlier at the regional level, men scholars in MENA are more likely to be last authors than their women colleagues in most countries. Few countries standout such as Egypt (2008, 2012 and 2016), Jordan (2008 and 2016), Turkey (2008), UAE (2016) and Pakistan (2016) which show no substantial difference in terms of probability of last authorship. As for the probabilities of first authorship, the majority of MENA countries do not show large differences.

We also notice the probability of being last author for women and men researchers is lower than the probability of being first author across all countries. Additionally, there was an overall shift in terms of probability levels when comparing the three cohorts. The 2008-cohort had a higher probability of last authorship than the 2012 and the 2016 cohort for most countries. When comparing the three cohorts, the differences in terms of last authorships seem to become smaller in UAE, Qatar, Pakistan, Iraq and Jordan.

In our study, we notice relatively large gender differences in the scientific workforce between countries of the same region. MENA countries have their own specifics and they have organized their research systems differently and researcher as a profession has also a different status. Some of the factors we analysed, such as more men starting a career as publishing researchers than women, and men producing on average more publications than women, cannot be explained by using bibliometric methods. Similar questions have been addressed in other studies (Ceci et al., 2014; Cheryan et al., 2017; Wang & Degol, 2017) and are not discussed in our study.

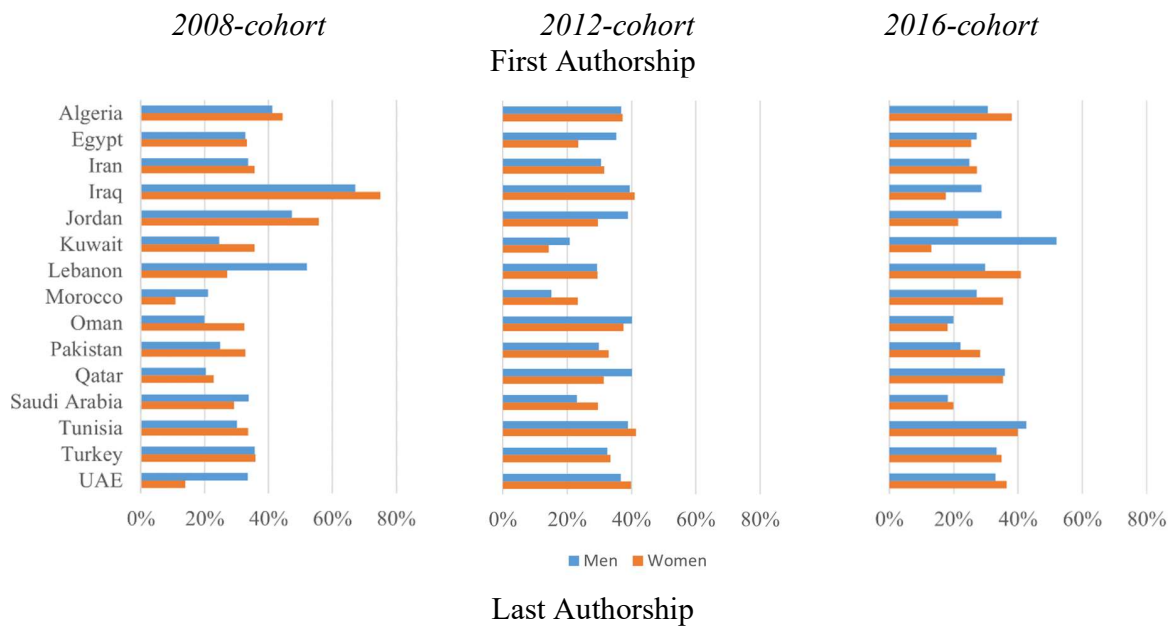

Last Authorship

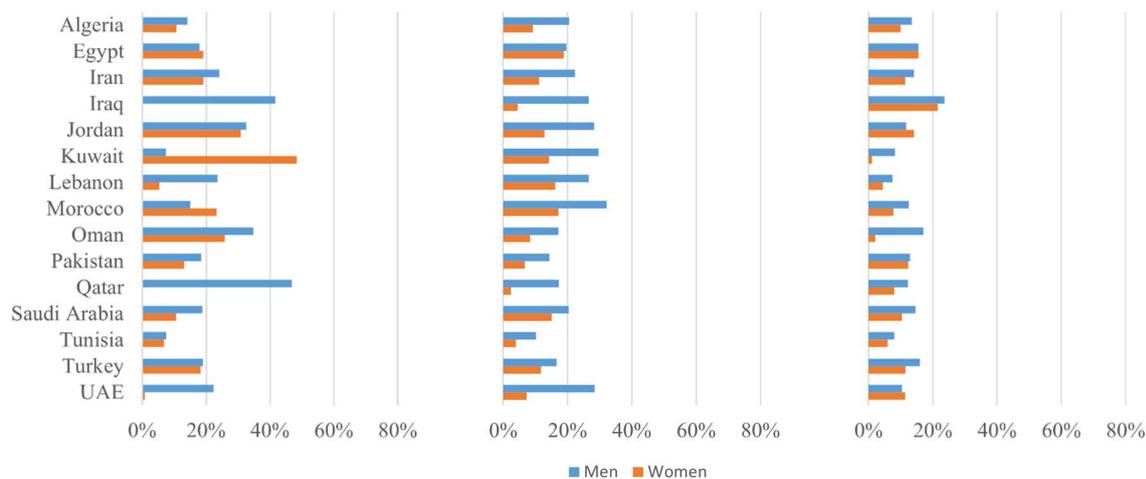

**Figure 11. Probability of being first author (top) or last (bottom) by country in year 5 of their career for men and women who started their career in 2008, 2012 and 2016.**

## 5. Discussion

Over the last few decades, the topic of women's empowerment in the Middle East and North Africa (MENA) region as well as in Islamic Societies (Offenhauer & Buchalter, 2005) has gained traction (OECD, 2014; Solati, 2017; UNDP, 2009; World Bank, 2011) In 2012, the World Bank chose gender equality and development as the main theme for its annual flagship publication (World Bank, 2012), which was followed by a special report focusing on the MENA region, reflecting the topic's growing importance and urgency since the region's rising political and social unrest (World Bank, 2013). We have contributed to this conversation by examining, nearly a decade after these initial publications, the state of gender in science in this region.

Our study provides a quantitative analysis of the gaps between men and women scientific authors in the MENA region on their representation, research productivity, and seniority. The results show that men scholars dominate in both number of authors and productivity of individual scholars. Disparities are reflected in every country of the region. In our study, we notice some MENA countries are close to gender parity in terms of participation of women in the science systems. Some of the gaps are relatively small in countries such as Tunisia, Lebanon, Turkey, Egypt, Iran, Morocco, and Algeria. When comparing the 2008 and 2016 cohorts, the percentage of women authors has increased by 7 percentage points, on average. Tunisia, Qatar, Jordan, and Iran have had the greatest increases. But we also observe large differences between countries in MENA. The proportion of women scholars in Iraq declined by 1%, while it increased only by 1% in Saudi Arabia, Kuwait, and Egypt. In terms of productivity, there was no significant difference between men and women in their first year of work. However, in the following years, men's output was between 11 and 51 percent higher than women's, with the gap widening over time. In the MENA region, men's probability of being the last author were 1.04 to 1.73 times higher than women's. In the first three to five years of a researcher's career, this probability reduced slightly. Then, it increased earlier for men than for women of the same cohorts, which suggests that men progressed to higher levels of seniority faster than women. We need to keep in mind that our results have some limitations as listed earlier. First, we focused exclusively on bibliometric parameters.

Researchers' activities that cannot be measured with bibliometric methods are not considered in our study. Then, in our gender inference approach based on first names, we used a binary perspective that ignores non-binary researchers. Without specific descriptions of the role of each author, such as promoted in the Contributor Roles Taxonomy (CRediT) initiative[3], it remains difficult to assess authors contributions. Finally. we studied only gender gap differences on few aspects of scientific research based on Web of Science Core Collection data. Since national languages are often generally used in fields where we have national applications, data from other citation indices with more regional content, such as the Arabic Citation Index, might also be used in future work to provide more comprehensive analysis for policy makers

Several possible reasons have been described to explain the evidence that men publish more than women during their career (Abramo, D'Angelo, et al., 2009; Larivière et al., 2011; Xie et al., 2003) such as differences in family responsibilities (Carr et al., 1998; Fox, 2005; Stack, 2004), academic rank (Van den Besselaar & Sandström, 2017), or career absence (Cameron et al., 2016). Some have argued that women's attrition in science is primarily attributed to women's personal motives and the notion of societal division of labour (Ceci & Williams, 2011; Ceci et al., 2009; Fox, 2005, 2006; Fox et al., 2017; Tasci, 2021). Why are there more men than women starting a career as a publishing researcher? And why do men publish on average more scientific papers than women? These questions have been widely discussed in other studies (Ceci et al., 2014; Cheryan et al., 2017; Wang & Degol, 2017) and these questions cannot be answered using bibliometric methods.

Karam and Afiouni (2014) call policy makers across the MENA region to adopt policies that support family burdens on women which would help them in their scientific careers and would result in a more balanced research ecosystem. Such policies are necessary to create a more inclusive development path. However, focusing simply on educational attainment may not be the most successful strategy for promoting women's empowerment in all levels in the region (Shalaby, 2014). Indeed, the rising educational attainment of women but with a low economic participation has been referred as the MENA paradox by the World Bank (2013). Assaad et al. (2020) argue the MENA paradox is largely due to changes in opportunity structures faced by educated women in the 2000s rather than to supply-side factors typically mentioned in the literature. To achieve true gender parity in MENA, several impediments on the societal, structural, institutional, and legal levels must be overcome simultaneously (Momani, 2016). Tasci (2021) has also provided a few recommendations to strengthen women scientists in the international research landscape. In the end, determining the best policies is very reliant on the context of each country. Rather than relying on previous experience, carefully organized small-scale trials of suggested initiatives are preferred before scaling them up to a national or regional level.

In terms of gender policies, MENA is catching up with other emerging regions, with Turkey, the UAE, Saudi Arabia, Morocco and Tunisia leading the region. However, MENA still lags behind all other regions of the world, with women in the region having around half the rights of men throughout all professional stages. This suggests, there is a strong need to improve the legal framework to give more equal opportunities regardless of gender. Previous studies have shown there is much to be gained by policies (Metcalfe, 2008). For example, in the UAE, Patterson et al. (2020) show that gender discrimination is on the decline, yet the problem still exists, necessitating efforts from policy makers, society, and governments to attain gender parity. Several efforts have been made by the Gulf Cooperation Council (GCC) states, particularly the United Arab Emirates and Saudi Arabia, which have significantly expanded their promotion of women in public life in recent years, with more women achieving high positions in a variety of social areas (Abdulkadir & Müller, 2020; Parveen, 2021; Rizvi & Hussain, 2021). In recent years, developments in the UAE and Saudi Arabia have enhanced legal equality for women. According to our research, the UAE has made improvement in terms of women's engagement in science, whereas Saudi Arabia still lags behind. Although North African nations are all lower income countries, they already show a higher level of women participation in science compared to GCC countries. It seems the UAE and Saudi Arabia's developments changes had also a ripple effect in other nations such as Jordan and Qatar where we already see significant growth in women's participation in science.

---

[3] https://credit.niso.org/

This study raises some important questions about the temporality of the policies and the temporality of our bibliometric analysis. It is difficult to analyse how the recent policies have potentially influenced the changes observed in our bibliometric study. We speculate that the observed changes in terms of policy development as well as the evolution perceived in bibliometrics result from underlying changes in the society. To a significant extent, societal and cultural changes are responsible for such changes. As of now, it is still too early to see changes in science systems of MENA nations which recently engaged with gender policies. However, the MENA region is catching up in terms of policy engagement and women representation in science. Based on our analysis, one can predict countries such as Turkey, Morocco, Algeria, Egypt, UAE, Qatar, Jordan and Iran might possibly close the women representation gap in science in the next 10 years. How soon the overall MENA region will close the gender gap in graduate studies as well as research is yet to be seen, but the recent progress is promising.

## 6. Acknowledgments

## 7. Author contributions

Jamal El-Ouahi: Conceptualization, Data curation, Formal analysis, Investigation, Methodology, Software, Visualization, Writing-original draft.

Vincent Larivière: Conceptualization, Formal analysis, Methodology, Writing-review & editing.

## 8. Funding and competing interests

The authors received no funding for this study. Jamal El-Ouahi is an employee of Clarivate Analytics, the provider of the Web of Science.

**Appendix A: Evaluation of our approach to gender inference**

To assess our algorithmic approach to gender inference, we used the validation data set created by Larivière et al. (2013). Out of the 3,704 authors in this data set, 667 were included in our study. These 667 authors were inferred a gender (women, men or unknown).

Table A lists the number and the corresponding percentage of correct and incorrect gender inferences. An incorrect result has been provided by our gender inference as follows: 6.1% of the authors were inferred a men gender and 10.1% of the authors were inferred a women gender. Table A also lists the results provided by the gender inference approach of Larivière et al. (2013) for the same set of 667 authors. The approach used by Larivière et al. (2013) has provided an incorrect result for 5% of the authors inferred to be men, and 15% of the authors inferred to be women. This means that, for authors inferred to be men, the approach of Larivière et al. (2013) is slightly more accurate than ours. However, for authors inferred to be women, their approach has a lower accuracy than ours.

**Table A. Statistics for our approach to infer a gender algorithmically and for the approach of Larivière et al. (2013).**

|  | Our approach | | | | Approach of Larivière et al. (2013) | | | |
|---|---|---|---|---|---|---|---|---|
|  | Men | | Women | | Men | | Women | |
| **Men** | 400 | 93.9% | 26 | 6.1% | 190 | 95.0% | 10 | 5.0% |
| **Women** | 17 | 10.1% | 151 | 89.9% | 23 | 15.0% | 130 | 85.0% |
| **Unknown** | 46 | 63.0% | 27 | 37.0% | 250 | 79.6% | 64 | 20.4% |